
\input phyzzx

\Pubnum{ \vbox{ \hbox{DAMTP-96-29} \hbox{9606025}} }
\pubtype{}
\date{June 1996}

\titlepage

\title{Killing spinors, the adS black hole and I\big(ISO(2,1)\big) gravity}

\author{R. Mann }
\address{Dept. of Physics
\break University of Waterloo
\break Waterloo, Ontario
\break Canada N2L 3G1}
\andauthor{G. Papadopoulos}
\address{D.A.M.T.P
 \break University of Cambridge\break
         Silver Street \break Cambridge CB3 9EW }

\abstract { We construct a supersymmetric extension of 
the $I\big(ISO(2,1)\big)$
Chern-Simons gravity and show that certain particle-like solutions and the
adS black-hole solution of this theory are supersymmetric.}

\endpage

\pagenumber=2

\def\pp{=\hskip-4.8pt \raise 0.7pt\hbox{${}_\vert$} ^{\phantom 7}}

\def\cM {{\cal M}}

\def\cL{{\cal{L}}}
\def\ch {{\cal L}(H)}

\font\mybb=msbm10 at 12pt
\def\bb#1{\hbox{\mybb#1}}
\def\bE {\bb{E}}
\def\la{\langle}
\def\ra{\rangle}

\REF\ta{A. Ach\' ucarro and P.K. Townsend, Phys. Lett. {\bf 180B} (1986) 89.}
\REF\witten{E. Witten, Nucl. Phys. {\bf B311} (1988) 46; {\bf B323} (1989) 113.}
\REF\amann{S. Carlip, J. Gegenberg and R.B. Mann, Phys. Rev. {\bf D51}
(1995) 6854; S. Carlip and J. Gegenberg, Phys. Rev. {\bf D44} (1991) 424.}
\REF\btz{M. Ba\~ nados, C. Teitelboim and J. Zanelli, Phys. Rev. Lett. {\bf 69} (1992) 1849.}
\REF\strom{K. Becker, M. Becker and A. Strominger, Phys. Rev D {\bf 51} (1995) 6603.}
\REF\pth{P.S. Howe, J.M. Izquierdo, G. Papadopoulos and P.K. Townsend,
Nucl. Phys.
{\bf B467} (1996) 183.}
\REF\itown{J.M. Izquierdo, and P.K.Townsend, Class. Quantum Grav. {\bf 12} (1995) 895.}
\REF\djh{S. Deser, R. Jackiw and G t'Hooft, Ann. Phys. {\bf 152} (1984) 220.}
\REF\dj {S. Deser and R. Jackiw, Ann. Phys. {\bf 153} (1984) 405.}
\REF\jol {J.D. Brown, J. Creighton and R.B. Mann, Phys. Rev. {\bf D50}
(1994) 6394.}
\REF\bcj{D. Bak, D. Cangemi and R. Jackiw, Phys. Rev. {bf D49} (1994) 5173.}
\REF\ms{G. Moore and N. Seiberg, Phys. Lett. {\bf 220B} (1989) 422.}
\REF\em{S. Elitzur, G. Moore, A. Schwimmer and N. Seiberg, Nucl. Phys. {\bf B326} (1989) 108.}


\chapter{Introduction}

Pure gravity in 2+1 dimensions has no propagating degrees of freedom and 
can be formulated as a  Chern-Simons (CS) gauge theory with gauge group
$ISO(2,1)$ [\ta].  As such, it is integrable and can be consistently 
quantised [\witten]. For a generic coupling of 2+1 gravity to matter, 
though, the above novel integrability properties of the theory are not retained.  However
there are couplings of 2+1 gravity to matter such that the combined 
gravity-plus-matter theory  again admits a Chern-Simons formulation.  
Such theories are also integrable and can be consistently quantised.  
There are several 2+1 gravity-matter systems that admit a CS formulation: 
for example all CS
theories with gauge group $G$ that have $ISO(2,1)$ as subgroup.  For the
purposes of this paper we will restrict ourselves to the model suggested in
[\amann];  this theory is  a Chern-Simons theory with gauge group
$I\big(ISO(2,1)\big)$.  The novel property of this theory is [\amann] that one of its
classical solutions is the adS black hole solution of [\btz];  this is the case
because it can be arranged such that topological matter can play the role of the
cosmological constant of the adS gravity.

The non-vanishing  commutators of the Lie algebra of
$I\big(ISO(2,1)\big)$ are 
$$
\eqalign{
[M_a,M_b]&=-\epsilon_{ab}{}^c M_c\ , \qquad [M_a, P_b]=-\epsilon_{ab}{}^c P_c
\cr
[M_a, J_b]&=-\epsilon_{ab}{}^c J_c\ ,\qquad [M_a, K_b]=-\epsilon_{ab}{}^c K_c
\cr
[J_a, K_b]&=-\epsilon_{ab}{}^c P_c\ ,}
\eqn\inone
$$
where $\{M_a; a=1,2,3\}$ are the Lorentz generators and $\{P_a; a=1,2,3\}$ are the momentum
generators associated with the gravitational sector, and $\{J_a; a=1,2,3\}$ and $\{K_a;
a=1,2,3\}$ are two sets of additional generators that are associated with the matter sector.
An invariant non-degenerate inner product is
$$
<M_a, P_b>=\mu\ \eta_{ab}\ , \qquad <J_a,K_b>=\mu\ \eta_{ab}\ ,
\eqn\intwo
$$
where $\mu$ is a non-zero real number. Given the algebra \inone\ and the
invariant non-degenerate inner product \intwo,  it is straightforward to
construct the associated Chern-Simons theory.  The fields of the theory are the
connections ${\cal A}$ with gauge group $I\big(ISO(2,1)\big)$ and the action is the
Chern-Simons functional of these connections.  This action describes the
coupling of Poincar\'e gravity to topological matter, is invariant under the
gauge transformations with gauge group $I\big(ISO(2,1)\big)$ up to surface terms, and the
field equations are the zero curvature conditions ${\cal F}({\cal A})=0$ of the
connection ${\cal A}$.  

Recently there has been some progress in finding the necessary and 
sufficient conditions for the existence of Killing spinors in 
2+1-dimensional spacetimes
for both Poincar\'e and adS gravities [\strom,\pth, \itown].  In the Poincar\'e case, the
existence of Killing spinors necessitates the generalisation of the standard
Poincar\'e supergravities [\pth].  The new Poincar\'e supergravities are
Chern-Simons theories whose Lie algebras have generators that are either
(i) those of the super-Poincar\'e algebra with central charges or
(ii) some of the outer automorphisms of the super-Poincar\'e algebra 
with central charges that rotate
its supersymmetry and central charges. 
The novelty here is that the gravitino
supersymmetry transformation of these new Poincar\'e supergravities is
modified by terms that include the gauge field associated with the
automorphism generators.  Such modification of the gravitino supersymmetry
transformations is necessary for the existence of the Killing spinors in these
theories [\pth]. 

The main purpose of this paper is to construct a supersymmetric extension of
2+1 gravity-matter system that is described by the $I\big(ISO(2,1)\big)$ Chern-Simons
theory above and to investigate the existence of Killing spinors for some
classical solutions of this theory.  For this a supersymmetric extension of the
algebra \inone\ will be constructed.  The generators of this supersymmetry
algebra will be the bosonic generators $\{ P_a, M_a, J_a, K_a\}$ of \inone, two
supersymmetry charges $\{Q^i; i=1,2\}$ and two additional bosonic charges
$\{Z,T\}$ where $Z$ is a central charge and $T$ is the generator of the outer
automorphisms. The gravitino supersymmetry transformation will determine the
Killing spinor equation and then the methods of [\pth] will be applied to
solve it for various solutions of the supergravity theory.  We will find that the adS black
hole solution of this theory admits a Killing spinor. 

This paper is organised as follows:   In section two we will
construct a supersymmetric extension of the $I\big(ISO(2,1)\big)$  Chern-Simons theory
and give the gravitino supersymmetry transformations.  In section three
we present several classical solutions of this theory and then we will
investigate the conditions under which such spacetimes admit Killing spinors.
Finally, in appendix one, we use the covariant canonical approach to
define the non-abelian charges of a CS theory and compute their Poisson
bracket algebra; this Poisson bracket algebra is a Kac-Moody algebra with a central
extension.


\chapter{Supergravity}

To construct a  supersymmetric extension of the the bosonic gravity-matter
system described by the $I\big(ISO(2,1)\big)$  Chern Simons theory [\amann], we 
must find a supersymmetric extension of the Lie algebra of 
$I\big(ISO(2,1)\big)$ together with an invariant non-degenerate inner 
product.   The non-vanishing commutators of such a superalgebra are 
as follows:
$$
\eqalign{
[M_a,M_b]&=-\epsilon_{ab}{}^c M_c\ , \qquad [M_a, P_b]=-\epsilon_{ab}{}^c P_c
\cr
[M_a, J_b]&=-\epsilon_{ab}{}^c J_c\ ,\qquad [M_a, K_b]=-\epsilon_{ab}{}^c K_c
\cr
[J_a, K_b]&=-\epsilon_{ab}{}^c P_c\ , \qquad a,b,c=1,2,3\ ,
\cr
\{ Q^i_\alpha,Q^j_\beta\} &=-{1\over2}\delta^{ij}  (\gamma^a)_{\alpha\beta} P_a
+i \epsilon_{\alpha\beta}\epsilon^{ij} Z
\cr
[M_a,Q_\alpha^i] &= {i\over2}(Q^i\gamma_a )_\alpha  , \qquad
[T,Q^i_\alpha] = -\epsilon^{ij}Q^j_\alpha\ ,}
\eqn\sone
$$
where $\{Q^i; i=1,2\}$ are the supersymmetry charges, $Z$ is a 
central charge, $T$ is an automorphism generator that rotates the 
two supersymmetry charges, and $\alpha, \beta=1,2$ are spinor indices. 
The rest of the generators are as those of the $I\big(ISO(2,1)\big)$ 
algebra of  \inone.  We use the `mostly-minus' metric convention and
hence gamma matrices that are pure imaginary. Note that
$$
\gamma^a\gamma^b =\eta^{ab} + i\epsilon^{abc}\gamma_c\ \qquad
(\bar\psi)_\alpha=\psi^\beta \epsilon_{\beta\alpha}.
\eqn\twob
$$
We also introduce a formal conjugation with respect to which all the 
even generators of the superalgebra are antihermitian whereas the odd 
ones are hermitian,  and we adopt the standard convention that complex 
conjugation of a fermion bi-linear introduces an additional minus sign.  

Next we introduce the folowing inner product
$$\eqalign{
\la M_a,P_b\ra &= \mu\, \eta_{ab} \qquad \la Q^i_\alpha,Q^j_\beta\ra =i \mu\,
\epsilon_{\alpha\beta}\delta^{ij}\ ,\cr 
\la T,Z\ra &=-\mu\ , \qquad <J_a,K_b>=\mu\,\eta_{ab}\ ,}
\eqn\twoc
$$
where $\mu$ is a real non-zero constant with dimensions of mass. This inner
product is invariant, non-degenarate and  hermitian
with respect to the formal conjugation introduced above. The
(anti-hermitian) gauge field ${\cal A}$ is
$$
{\cal A}= e^aP_a + \omega^aM_a + CZ +AT +B^aJ_a+H^a K_a+ \psi^iQ_i\ ,
\eqn\twoc
$$
where $e^a$ is interpreted as the frame, $\omega^a$ is the 
spin-connection of the gravitational sector, $\{\psi^i; i=1,2\}$ are 
the gravitinos,  and the rest $\{ C, A, B^a, H^a\}$ are matter gauge fields  
associated with the remaining generators of the superalgebra \sone. 
\foot{Note that in our conventions the field
$\psi$ anticommutes with $Q$.} We compute the curvature two-form
${\cal F}=d{\cal A}+{\cal A}^2$ to be
$$
{\cal F}= T^aP_a + F^a(\omega)M_a +F(C)Z + F(A) T + F^a(B)J_a+F^a(H)K_a+{\cal
D}\psi^i Q_i\ ,
\eqn\twod
$$
where
$$
\eqalign{
T^a &= de^a - \epsilon^a{}_{bc} \omega^b e^c
-{1\over4}\bar\psi^i\gamma^a\psi^i-\epsilon^a{}_{bc} B^b H^c
\cr
 F^a(\omega) &=d\omega^a -{1\over2}\epsilon^a{}_{bc}\omega^b\omega^c\cr
F(C) &=dC\cr
F(A) &=dA \cr
F^a(B)&=dB^a-\epsilon^a{}_{bc}\omega^b B^c
\cr
F^a(H)&=dH^a-\epsilon^a{}_{bc}\omega^b H^c 
\cr
{\cal D}\psi^i  &= d\psi^i +{i\over2} \omega^c\gamma_c\psi^i +
A\epsilon^{ij}\psi^j\ . }
\eqn\twoee
$$
Note that $\omega^a$  has torsion proportional not only to gravitinos, 
as is expected in a supergravity theory, but also proportional to 
other matter fields like $B$ and $H$.  

The action of the theory is
$$\eqalign{
S=&\mu\int d^3x\ \big[  e R(\omega) -i  \varepsilon^{mnp}\bar\psi^i_m{\cal
D}_n\psi^i_p +2  \varepsilon^{mnp}C_m\partial_n A_p 
\cr
&+2  \varepsilon^{mnp}
H^a_m\partial_n B^a_p-2 \epsilon_{abc}\varepsilon^{mnp} \omega^a_m B_n^b
H_p^c\big]\ ,}
\eqn\twod
$$
where the `Ricci' scalar $R(\omega)$ is the trace of the frame $e$ with Hodge dual tensor of the
curvature $F(\omega)$  and   the spin-connection
$\omega_m{}^a$ is an independent  field, i.e. this is the first-order form of the
supergravity action. The action \twod\ is invariant (up to a surface term) under gauge
transformations of the connection ${\cal A}$. The supersymmetry 
transformations are the gauge transformations along the 
fermionic generators $Q^i$ of the supersymmetry algebra.  The 
non-zero supersymmetry transformations of the fields are
$$
\eqalign{
\delta e^a &= {1\over 2}\bar\zeta^i\gamma^a \psi^i\cr
\delta \psi^i &= {\cal D}\zeta^i\cr
\delta C &= i\epsilon_{ij}\bar\zeta^i\psi^j\ , }
\eqn\twoe
$$
where $\zeta^i$ are anticommuting spinor parameters and
$$
{\cal D}\zeta^i=d \zeta^i+ {i\over2} \omega^c \gamma_c\zeta^i +
A\epsilon^{ij}\zeta^j\  .
\eqn\sfour
$$
Note that the non-vanishing supersymmetry transformations of the fields are
precisely those of the Poincar\'e supergravity theories of [\pth].


\chapter{ Killing Spinors}

\section{Particle-like solutions}

A class of {\it stationary} classical solutions of the 2+1 
gravity-matter system of the previous
section is given by generalising the multi-point particle solutions of the
2+1 Poincar\'e gravity of [\djh, \dj]. 
For this, we use the ansatz
$$\eqalign{
ds^2=& \left(dt +\sum_{L=1}^N J_L(r)
{{\vec{r}-\vec{r}_L}\over{|\vec{r}-\vec{r}_L|^2}}\times d\vec{r}
\right)^2-h^2(r) d\vec{r}\cdot d\vec{r}, 
\qquad \omega^{\underline i}=0, \qquad H^{\underline 0}=0,
\cr
B^{\underline i}=& \epsilon^{\underline i}{}_{\underline j} H^{\underline j}\ ,
\qquad H^{\underline i}=\phi dx^i\ ,}
\eqn\ckone
$$
where $\phi, h$ are functions of the Euclidean 2-space, $\bE^2$,  whose
coordinates are $\vec{r} = (x,y)$, the underline denotes
frame indices,  and $\epsilon$  is
the $\epsilon$-tensor on {\it Euclidean} 2-space $\bE^2$, {\rm i.e.}  $\epsilon{\underline
 i}{}_{\underline j}=\delta^{\underline i \underline k} \epsilon_{\underline k \underline j}$.

Choosing the triad so that
$$
\eqalign{
e^{\underline 0} &= dt + \sum_{L=1}^N J_L(r)
{{\vec{r}-\vec{r}_L}\over{|\vec{r}-\vec{r}_L|^2}}\times d\vec{r} \cr
e^{\underline i} &= h\ dx^i\ ,}
\eqn\ckoneaa
$$
it is straightforward to show that $de^{\underline 0} = 0$ provided
the functions $J_L$ are all constants.  Incorporating this into
the above ansatz, the field equations
$F(B)=0$ and $F(H)=0$ yield
$$
dB^{\underline 0}=0\ , \qquad \omega^{\underline 0}_i=\epsilon^{\underline j}{}_{\underline i}
\partial_j \log\phi\
\eqn\ckonea
$$
so 
$$
B^{\underline 0}=df\ .
\eqn\ckoneab
$$ 
{}From the field equations $F(\omega)=0$, 
we find that $\log\phi$ is a {\it harmonic} function on $\bE^2$, so we can write  
$$
\phi^2=\Pi_{L=1}^N |\vec{r}-\vec{r}_L|^{-{\rho_L\over \pi}}\ ,
\eqn\ckonec
$$
where  $\{\rho_L;L=1,\dots, N\}$ are real constants and $\{\vec{r}_L; L=1,\dots n\}$ are the
centres of the harmonic functions. Next, the field equation 
$T=0$ implies that 
$$
f={h\over \phi}\  , 
\eqn\ckoneb
$$
and $\log f$ is a {\it harmonic} function on $\bE^2$. 
Finally the  field equations for $A$ and $C$ decouple 
and any flat connection solves their field equations. 

It is clear from the above that since $\log \phi$ and $\log h$ are harmonic functions on
$\bE^2$,
$\log h$ is a { \it harmonic}  function on $\bE^2$ as well, so we may choose
$$
h^2=\Pi_{L=1}^N |\vec{r}-\vec{r}_L|^{-{m_L\over \pi}}\ , 
\eqn\ckoned
$$
in which case  the metric \ckone\ is that of a conical N-particle spacetime located at the
positions $\{\vec{r}_L; L=1,\dots N\}$ where 
$\{m_L; L=1,\dots, N\}$ are the masses of the particles, and
$\{J_L; L=1,\dots, N\}$ are their spins. To explain this identification of the 
mass and spin of the $L$th-particle with the parameters $m_L$ 
and $J_L$ of the metric \ckone, respectively, we remark that the mass and 
the spin of the $L$th-particle are
determined in terms of the geometry of the configuration as follows:
$$
m_L\equiv \oint_{\Gamma_L} \Omega^{\underline 0}\ ,\qquad J_L\equiv {1\over 2\pi}\oint_{\Gamma_L}
e^{\underline 0}\ ,
\eqn\ckoneda
$$
where $\Omega$ is the Levi-Civita connection of the metric \ckone{} and
$\Gamma_L$ is a
non-trivial fundamental homology cycle of the space\foot{The fundamental homology cycles are the
closed paths in space that their interior contain the position of only one particle.}. The
total mass of the configuration is the deficit angle of the conical spacetime and therefore it
should be less than
$2\pi$. Finally,  the function $f$ is easily computed using \ckonec, \ckoneb and \ckoned as
follows:
$$
f^2=\Pi_{L=1}^N |\vec{r}-\vec{r}_L|^{-{(m_L-\rho_L)\over \pi}}_L\ .
\eqn\cktwo
$$
 
The gravitino supersymmetry transformation is expressed in terms of the 
connections $\omega$ and $A$. For the solutions above both $\omega$ and $A$ connections are flat, so a Killing
spinor exists (after an appropriate projection) for the subclass of the above solutions that the
holonomies of the connections $\omega$ and $A$ obey the holonomy matching condition of [\pth]. 
This condition can be expressed in terms of the charges (holonomies) of the
connections $\omega$ and $A$ evaluated at the non-trivial fundamental homology 
cycles $\{\Gamma_L; L=1,\dots,N\}$ of 
spacetime.  The holonomy of the connection
$\omega$ evaluated at the fundamental cycles is given by the set of  numbers,
$\{\rho_L;L=1,\dots, N\}$, i.e.
$$
\rho_L=\oint_{\Gamma_L} \omega^{\underline 0} \ .
\eqn\ckfive
$$
Using large gauge transformations, we can restrict these number as follows:
$0\leq \rho_L < 2\pi$.  Similarly the holonomy of the connection $A$ is
$$
Q_L=\oint_{\Gamma_L} A\ .
\eqn\cksix
$$ 
Then the holonomy matching conditions are given as follows:
$$
\rho_L =2 |Q_L|
\eqn\ckthree
$$
for periodic boundary conditions for Killing spinors on the holonomy
fundamental cycles (even spin structure), and
$$
\rho_L= 2 |Q_L\pm\pi|
\eqn\ckfour
$$
for antiperiodic boundary conditions for Killing spinors on the holonomy
fundamental cycles (odd spin structure).

\section{The adS black hole solution}

Another solution of interest of this supergravity plus matter theory is the adS
black hole [\btz].  This can be easily seen by taking the gravitino equal to
zero and then observing that the field equations for gauge fields  $A$ and $C$
decouple from the field equations of the remainning fields.
This black hole solution is given as follows
$$
\eqalign{ ds^2&=-N^2 dt^2+N^{-2} dr^2+r^2 (Wdt+d\phi)^2
\cr
B^0&= r_+d\phi+{r_-\over \ell}dt\ , \quad B^1=-\ell d(\nu+{\sqrt{\nu^2-1}})\ ,
\quad B^2={r_+\over \ell} dt-r_- d\phi
\cr
H^0&=-{1\over \ell} B^0\ , \quad H^1= d(\nu-{\sqrt{\nu^2-1}})\ ,\quad
H^2={1\over \ell} B^2\ ,}
\eqn\ckseven
$$
where
$$\eqalign{
N^2(r)&=-M+{r^2\over \ell^2}+{J^2\over 4r^2}\ , \quad W=-{J\over 2r^2}
\cr
\nu^2&={r^2-r^2_-\over r_+^2-r^2_-}\ ,}
\eqn\ckeight
$$
$-\infty <t<\infty$, $0\phi<2\pi$ and $0<r<+\infty$.

The parameters $\ell$, $M$ and $J$ are constants of integration
in \ckseven. For
the usual adS black hole [\btz], $\ell$ is a coupling constant 
inversely proportional to the square root of the cosmological constant,
and $M$ and $J$ are the quasilocal mass and angular momentum of
the black hole [\jol]. However for the gravity-plus-matter solution
\ckseven, \ckeight{} given above, it
has been shown that $M\ell$ is the angular momentum parameter 
and $J/\ell$ is the mass parameter [\amann].

To determine whether or not this solution admits Killing spinors
observe that the connection $\omega$ of the theory is identically zero.  So it
is flat and has zero holonomy for every closed path in space-time.  So the black
hole space-time admits Killing spinors provided that we take $A$ to be gauge
equivalent to a trivial $U(1)$ connection.


\chapter{Conclusion}

We have investigated the existence of killing spinors in the CS theory with
gauge group $I\big(ISO(2,1)\big)$ suggested in [\amann].  This theory
describes Poincar\'e
gravity coupled to topological matter and one of its classical solutions is the
adS black hole.  For this we have constructed a supersymmetric extension of this
theory using a super-extension of $I\big(ISO(2,1)\big)$ and CS methods.  We have then
shown that some of the classical solutions of this theory are supersymmetric,
i.e. they admit Killing spinors.  Amongst the supersymmetric solutions are
spinning particle-like configurations and the 
adS black hole solution of this theory. 

\vskip 0.5cm

{\bf Acknowledgements:}
This work was supported in part by the Natural Sciences and
Engineering Research Council of Canada. G.P. is supported by a Royal Society University
Research fellowship.

\appendix

\centerline{\bf Non-abelian charges}
In this appendix, we will summarize the properties of the currents of a CS
theory using the covariant canonical approach.  This will complement the
relevant part of section three.   Let
$\cL(\phi)$ be a Lagrangian at most quadratic in the time derivatives of the
fields.  Moreover let us suppose that
$\cL(\phi)$ is invariant under the transformations
$$
\delta_u\phi=f(u,\phi), \qquad [\delta_u, \delta_v]\phi=
\delta_{[u,v]}\phi\equiv f([u,v],\phi) 
\eqn\cone
$$
with parameters $u, v$ up to a surface term, i.e.
$$
\delta_u\cL(\phi)=\partial_\mu X^\mu(u, \phi)\ ,
\eqn\ctwo
$$
where $\mu$ is a spacetime index.
The field equations of the theory are
$$
{\partial\cL\over \partial\phi}-\partial_\mu\pi^\mu=0
\eqn\cthree
$$
where $\pi^\mu={\partial\cL\over
\partial\partial_\mu\phi\ }$.
One can define a symplectic current in the theory
$$
S^\mu=d\phi\wedge d\pi^\mu
\eqn\cfour
$$
which is conserved subject to field equations and closed as a form on the space
of fields of the theory.  Assuming that the spacetime is globally hyperbolic,  one
can define a closed two-form 
$$
\Omega=\int_{\Sigma} S^\mu d\Sigma_\mu
\eqn\cfive
$$
where $\Sigma$ is a Cauchy surface.  Due to the conservation of the symplectic
current $\Omega$ is independent from the choice of $\Sigma$.  If the
Lagrangian is not invariant under local (gauge) transformations then the form
$\Omega$ is non-degenerate and therefore symplectic (at least weakly).

The conserved currents, subject to field equations, of the theory associated with
the symmetries
\cone\ are
$$
J^\mu(u, \phi)=f(u,\phi) {\partial\cL\over\partial\partial_\mu\phi}- X^\mu(u,
\phi)\ .
\eqn\csix
$$
The conserved charges are
$$
Q[u, \phi]=\int_{\Sigma} J^\mu(u, \phi) d\Sigma_\mu\ .
\eqn\cseven
$$
The Poisson bracket of two of the above charges in the context of
the covariant canonical approach is defined as follows:
$$
\{Q[u, \phi], Q[v, \phi]\}\equiv\delta_u Q[v, \phi]-\delta_v Q[u, \phi]
\eqn\ceight
$$
Due to the surface terms in \csix\ that are necessary for the definition of
conserved charges, the Poisson algebra of the charges need not be isomorphic to
the Lie algebra of the transformations on the fields.

Next we apply this general formalism to the case of Chern-Simons theory in 2+1
dimensions; see also [\bcj].  To define a Chern-Simon theory in 2+1 dimensions, we
introduce a Lie algebra $\ch$ of a group $H$, an invariant non-degenerate
inner product $<\cdot,\cdot>$ on $\ch$ and gauge fields $A$
with gauge group $H$.  The  action is
$$
S=\mu\int_\cM (<A, dA>+{1\over3} <A, [A,A]>)\ ,
\eqn\cten
$$
where $\mu$ is a constant.
The symplectic current is
$$
S=\mu\ \, {}^*<dA, dA>
\eqn\ctena
$$
and the closed two-form $\Omega$ is
$$
\Omega=\mu\int_\Sigma \big({}^*<dA, dA>\big)_\mu d\Sigma^\mu\ ,
\eqn\ctenb
$$
where star is the Hodge star.
The form $\Omega$ vanishes, subject to field equations, along the directions
tangent to the gauge orbits and therefore is degenerate\foot{An alternative
way to use \ctenb\ is to view it as a symplectic form on the space of
connections of a two-dimensional Riemann surface.  Evaluating \ctenb\ along the
vectors tangent to  orbits of the gauge group, one gets as a momentum map the
curvature of the connection $A$. Using Hamiltonian reduction, one can define a
symplectic structure on the space of gauge equivalence classes of flat
connections.}.  The theory is invariant under, up to a surface term, under the
gauge transformations 
$$
\delta A=Du\equiv du+[A,u]
\eqn\celeven
$$
of the connection $A$ with parameter $u$.  The conserved currents are
$$
J=\mu\ \, {}^* d<u,A>\ .
\eqn\ctwelve
$$
The associated charges are
$$
Q[u,A]= \int_{\Sigma} J^\mu(a, \phi) d\Sigma_\mu=\mu\int_{\partial \Sigma} <a,A>\
,
\eqn\cthirteen
$$
where $\partial \Sigma$ is the boundary of $\Sigma$.
The associated Poisson brackets are
$$
\{Q[u,A], Q[v,A]\}=-2 Q\big[[u,v],A\big]-2 \mu\int_{\partial \Sigma} <u,dv>\ .
\eqn\cfourteen
$$
In the case that $\Sigma$ is a disc, then $\partial\Sigma=S^1$  and the Poisson bracket algebra
of charges in this case is isomorphic to a Kac-Moody algebra.  In addition, 
using the field equations
$F(A)=0$, one can write $A=h^{-1}dh$ in which case \cfourteen\ becomes the Poisson bracket
algebra of the left invariant currents of a Wess-Zumino-Witten model whose
target space is the group
$H$  in agreement with the results of [\ms, \em].

\refout

\end